\documentclass[pdflatex,sn-mathphys-num]{sn-jnl}

\usepackage{graphicx}
\usepackage{multirow}
\usepackage{amsmath,amssymb,amsfonts}
\usepackage{amsthm}
\usepackage[title]{appendix}
\usepackage{xcolor}
\usepackage{textcomp}
\usepackage{manyfoot}
\usepackage{booktabs}
\usepackage{algorithm}
\usepackage{algorithmicx}
\usepackage{algpseudocode}
\usepackage{listings}

\theoremstyle{thmstyleone}
\newtheorem{theorem}{Theorem}

\theoremstyle{thmstyletwo}
\newtheorem{example}{Example}

\theoremstyle{thmstylethree}
\newtheorem{definition}{Definition}

\begin{document}

\title[Frequency Estimation from the Internal State of the Boyer--Moore Majority Vote Algorithm]{Frequency Estimation from the Internal State of the Boyer–Moore Majority Vote Algorithm}

\author[1]{\fnm{Saeed} \sur{Momeni}}\email{smomeni@iasbs.ac.com}

\author*[1]{\fnm{Mohsen} \sur{Hooshmand}}\email{mohsen.hooshmand@iasbs.ac.ir}

\affil[1]{\orgname{Institute for Advanced Studies in Basic Sciences (IASBS)}, \orgaddress{\city{Zanjan}, \country{Iran}}}

\abstract{The Boyer-Moore Majority Vote algorithm identifies a majority candidate in a single pass over a data stream using constant memory $O(1)$ and linear processing time $O(n)$. While the classic algorithm identifies a candidate without directly estimating its frequency, its internal register states retain information about item occurrences during execution. This work investigates a frequency estimator derived directly from these internal counters. We establish the conditions under which the counter state enables exact recovery of the frequency of the majority element and characterize the estimation error when these conditions are not satisfied. In particular, we analyze the effect of stream ordering on the estimator and identify delayed-arrival configurations, such as Majority-Delayed sequences, that can lead to systematic estimation errors. We further derive the error behavior associated with late-arriving majority blocks and empirically evaluate the estimator across multiple stream topologies. The results demonstrate that the internal state of the Boyer-Moore algorithm can provide frequency information without increasing its $O(1)$ space complexity. At the same time, its accuracy depends critically on the ordering of elements in the input stream.}

\keywords{Data Streams, Boyer-Moore Algorithm, Frequency Estimation, Streaming Invariants, Concept Drift, Branch Misprediction}

\maketitle

\section{Introduction}\label{sec1}
Processing continuous data streams requires a careful balance between memory allocations and execution speed \cite{manku2002approximate, cormode2008finding, cormode2018frequent}. Many real-world systems, from network routing tables to distributed logging pipelines, must identify frequent items under strict hardware constraints \cite{demaine2002frequency, metwally2005efficient, benbasat2020frequency}. Formally, when given an input sequence $\sigma = \langle x_1, x_2, \dots, x_n \rangle$ made of items from a universe $\mathcal{U}$, the majority detection task looks for an item $M \in \mathcal{U}$ whose true frequency satisfies $f(M) > \lfloor n/2 \rfloor$. Storing full lookup tables to track every distinct key requires $\Omega(|\mathcal{U}|)$ memory, which quickly becomes impractical on embedded devices or high-speed network switches \cite{alon1999space, karp2003simple, li2021space}.

To operate within tight memory bounds, probabilistic synopses, sketch-based summaries, and approximate tracking schemes are widely used \cite{charikar2002finding, cormode2005count, wang2019sketch}. For instance, Lossy Counting by Manku and Motwani \cite{manku2002approximate} offers $\epsilon$-deficient counts over continuous streams with small memory footprints, whereas modern counter-based surveys highlight trade-offs between update throughput and per-item estimation fidelity \cite{benbasat2020frequency, cormode2018frequent}. Alongside these summary methods, the Boyer-Moore majority vote algorithm \cite{boyer1981fast} remains a classic deterministic choice. Its generalization to $k$-frequent items was formalized by Misra and Gries \cite{misra1982finding}, its average-case step counts were modeled by Tsai using Markov chains \cite{tsai2006average}, and Chassaing examined its optimal candidate tracking properties using martingales when elements arrive with unequal probabilities \cite{chassaing1997determining, chassaing1999average}. Recent advances in space-efficient streaming emphasize minimizing counter updates to maximize hardware cache locality and processing speed \cite{li2021space, wang2019sketch}.

However, standard Boyer-Moore implementations only output a single majority candidate without reporting how many times that item appeared \cite{boyer1981fast}. To explore whether we can estimate frequencies without adding memory allocations \cite{cormode2008finding}, our research focuses on tracking the internal register changes during algorithm execution. Specifically, we investigate the relationship between cancellation counts and terminal counter values in the algorithm's state machine, while analyzing hardware branch misprediction penalties \cite{smith1998study, fog2016microarchitecture}.
Existing work analyses the candidate-selection behavior of Boyer–Moore, whereas we investigate whether its internal execution state can itself be used as a frequency estimator and characterize the dependence of this estimator on stream ordering.

This work makes three primary contributions:
\begin{enumerate}
\item We define the internal counter relationship and provide an inductive proof demonstrating the conditions under which it equals the exact item frequency. 
\item We detail the boundary conditions that lead to the wrong estimation of frequencies by this estimator and provide an explicit formulation for tracking errors caused by late-arriving majority distributions.
\item We empirically analyze the behavior of the estimator across six stream paradigms, highlighting the significant linear scaling error observed under Majority-Delayed topologies, as well as the micro-architectural CPU pipeline penalties.
\end{enumerate}

\section{Algorithmic Instrumentation Logic}\label{sec2}
To monitor the evolution of internal variables during execution, we modified the standard Boyer-Moore loop to log key register events. Algorithm~\ref{alg:boyer_moore} presents this instrumented routine, which tracks candidate reassignments alongside every decrement step.

\begin{algorithm}[H]
\caption{Instrumented Boyer-Moore Stream Processor}\label{alg:boyer_moore}
\begin{algorithmic}[1]
\Require Stream sequence $\sigma = \langle x_1, x_2, \dots, x_n \rangle$
\State $C_n \leftarrow \phi$
\State $K \leftarrow 0$
\State $D_n \leftarrow 0$
\For{each $x_i \in \sigma$}
    \If{$K == 0$}
        \State $C_n \leftarrow x_i$
        \State $K \leftarrow 1$
    \ElsIf{$x_i == C_n$}
        \State $K \leftarrow K + 1$
    \Else
        \State $K \leftarrow K - 1$
        \State $D_n \leftarrow D_n + 1$
    \EndIf
\EndFor
\State $K_n \leftarrow K$
\State \Return $C_n, K_n, D_n$
\end{algorithmic}
\end{algorithm}

Most of the algorithm remains consistent with the original Boyer-Moore approach. While the original version returns only the candidate and its count, the updated version in Algorithm~\ref{alg:boyer_moore} returns an additional parameter, which we call ``$D_n$.'' As shown in Algorithm~\ref{alg:boyer_moore}, line 11 updates $D_n$ whenever an incoming stream element conflicts with the current candidate. This counter accumulates each cancellation event across the entire stream pass.

A subtle but important implementation detail of the instrumented routine in Algorithm~1 is the retention policy of the candidate register $C_n$ when the counter $K$ reaches zero. In the standard Boyer--Moore formulation, the candidate variable is often treated as undefined or reset when $K=0$. However, in our instrumented version, we deliberately keep the current value of $C_n$ unchanged even after a cancellation brings $K$ to zero. The register $C_n$ is only overwritten when a new element arrives while $K$ is already zero, triggering the candidate election branch (lines 5--7). This design choice does not affect the correctness of the majority vote algorithm, since a counter value of zero indicates that the stored candidate has no active voting weight; the value retained in $C_n$ is effectively dormant until it is either re-elected by a matching element or replaced by a new candidate. Nevertheless, this retention policy is critical for interpreting the tracing tables in Examples~\ref{ex:1} and~\ref{ex:2}, where $C_n$ may hold a non-$M$ value even after the counter has been fully depleted.

\section{Data Stream Topologies}\label{sec3}
To evaluate the mathematical resilience and boundary constraints of our internal counter tracking, we establish six distinct stream generation models:
\begin{itemize}
    \item \textbf{Uniform Distribution (\texttt{uniform}):} Randomly shuffles majority and minority keys throughout the stream length, assigning the majority item approximately equal probability of appearing at each stream position.
    \item \textbf{Long Runs Topology (\texttt{long\_runs}):} Groups majority items into continuous blocks between $\lfloor n/50 \rfloor$ and $\lfloor n/10 \rfloor$ elements long, separated by single minority entries.
    \item \textbf{Majority First (\texttt{majority\_first}):} Places all majority items in the front portion of the stream, with minority items trailing afterward.
    \[\sigma = \underbrace{M, \ldots, M}_{f(M)}, \underbrace{X_1, \ldots, X_{n-f(M)}}_{minority}\]
    \item \textbf{Majority-Delayed (\texttt{majority\_delayed}):} Positions the minority items first, placing all majority elements into the final stream block. This creates the exact worst-case failure mode analyzed in Theorem~\ref{thm:mismatch}.
    \[\sigma = \underbrace{X_1, \ldots, X_{n-f(M)}}_{minority}, \underbrace{M, \ldots, M}_{f(M)}\]
    \item \textbf{Alternating Sequence (\texttt{alternating}):} Interleaves majority and minority items in a strict round-robin sequence until minority items are exhausted, padding the remaining tail with majority elements.
    \item \textbf{Skewed Minority Distribution (\texttt{skewed}):} Structures majority items into blocks while concentrating 80\% of all minority instances into a single dominant key, creating heavy internal competition.
\end{itemize}

\section{Theoretical Counter Stability Analysis}\label{sec4}
Before examining the theoretical limits, it is critical to evaluate how the six data stream models defined in Section~III behave under our instrumented framework. We prove that the invariant property holds strictly for structured and convergent topologies (\texttt{long\_runs}, \texttt{majority\_first}, and \texttt{alternating}), while it collapses deterministically for delayed arrival topologies such as \texttt{majority\_delayed}. The underlying reason for this divergence is candidate register inertia: when majority elements appear early or are uniformly distributed, cancellations correspond directly to valid majority–minority pairs; conversely, when the majority block is delayed until the stream suffix, the majority elements act solely as counter-decrements against a false incumbent key, producing an unrecoverable negative estimation offset.

\begin{definition}[Internal-counter estimator]\label{def:estimator}
By having input stream $\sigma_i = \langle x_1, \dots, x_n \rangle$, internal-counter estimator $\hat{f}$ is:
\begin{equation}
\hat{f}_n = D_n + K_n,
\end{equation}
where $D_n$ is the total number of decrement operations during the complete execution, $K_n$ is the final value of the Boyer–Moore counter, and $C_n$ is the final candidate.
\end{definition}

\begin{theorem}[Exact recovery under correct candidate identification]\label{thm:convergence}
Let $M$ be the true majority element in stream $\sigma$. If the candidate register converges to $M$ by the end of execution ($C_n = M$), the estimation error $E_n = f(M) - \hat{f}_n$ equals zero:
\begin{equation}
E_n = f(M) - \hat{f}_n = 0
\end{equation}
\end{theorem}

\begin{proof}
For a strict majority M, let $f(M)=m$ and $n-m$ be the number of non-majority elements. 

If $C_n=M$, then by the pigeonhole principle, the algorithm must have canceled every non-$M$ element against an $M$ element. Each cancellation consumes exactly one $M$, so $D_n= n-m$. The remaining unmatched $M$ elements are $K_n=m-(n-m)=2m-n.$

Consequently, 
\begin{equation}
D_n+K_n=(n-m)+(2m-n)=m=f(M). 
\end{equation}
Thus, $\hat{f}_n=D_n+K_n=f(M)$ whenever the final candidate is the true majority.
\end{proof}

\begin{example}[Exact Frequency Invariance Under Convergence]\label{ex:1}
To illustrate the exact stability guaranteed by Theorem~\ref{thm:convergence}, consider the stream sequence of length $n=10$:
\[
\sigma = \langle 4, 4, 2, 1, 4, 3, 4, 1, 4, 4 \rangle
\]
where the majority target is $M=4$ with ground-truth cardinality $f(4) = 6$.
\begin{table}[h]
\centering
\caption{Tracing Algorithm~\ref{alg:boyer_moore} step-by-step over stream $ \sigma = \langle 4, 4, 2, 1, 4, 3, 4, 1, 4, 4 \rangle$ of Ex.\ref{ex:1}.}\label{tab:ex1}
\begin{tabular}{rrrrrrrrrrrr}
\toprule
$i$   & 0      & 1 & 2 & 3 & 4 & 5 & 6 & 7 & 8 & 9 & 10 \\ \midrule
$x_i$ & 0      & 4 & 4 & 2 & 1 & 4 & 3 & 4 & 1 & 4 & 4  \\ 
$K_i$ & 0      & 1 & 2 & 1 & 0 & 1 & 0 & 1 & 0 & 1 & 2  \\ 
$D_i$ & 0      & 0 & 0 & 1 & 2 & 2 & 3 & 3 & 4 & 4 & 4  \\ 
$C_i$ & $\phi$ & 4 & 4 & 4 & 4 & 4 & 4 & 4 & 4 & 4 & 4  \\ \bottomrule
\end{tabular}
\end{table}

Tracing Algorithm~\ref{alg:boyer_moore} step-by-step across this interleaved stream (Table~\ref{tab:ex1}):
\begin{enumerate}
    \item \textbf{Prefix reinforcement and decay ($i=1 \dots 4$):} The processor initializes with candidate $4$ ($K=2$). Incoming minority keys $2$ and $1$ decrement the active counter, yielding $K_4 = 0$ and $D_4 = 2$.
    \item \textbf{Interleaved re-election phases ($i=5 \dots 8$):} At $i=5$, element $4$ re-elects candidate $4$ ($K=1$), which is immediately canceled by minority item $3$ at $i=6$ ($K_6 = 0, D_6 = 3$). Similarly, $x_7=4$ restores candidate $4$ before $x_8=1$ cancels it, leaving $K_8 = 0$ and $D_8 = 4$.
    \item \textbf{Terminal convergence ($i=9, 10$):} The final two elements $x_9=4$ and $x_{10}=4$ re-elect and increment the register, completing execution with $C_{10} = 4$, $K_n = 2$, and $D_n = 4$.
\end{enumerate}
Evaluating the internal frequency estimator yields:
\[
\hat{F} = D_n + K_n = 4 + 2 = 6
\]
Computing the error difference metric verifies exact convergence:
\[
E_n = f(4) - \hat{f} = 6 - 6 = 0
\]
Because the candidate register successfully converged to $M=4$ at the end-of-file (EOF) ($C_n = M$), every cancellation directly mapped to an occurrence of the majority item, preserving Def.\ref{def:estimator} without estimation error.
\end{example}

\begin{theorem}[Estimation Error Under Candidate Mismatch]\label{thm:mismatch}
Let $M$ be a target item in a stream $\sigma$ of length $n$, and let
$C_n$ and $K_n$ denote the final candidate and final counter of the
Boyer--Moore algorithm, respectively. Define the estimator

\[
\hat f_n = D_n + K_n,
\]

where $D_n$ is the total number of decrement operations.

If the target item is not the final candidate, i.e., $C_n \neq M$,
then the estimation error

\[
E_n = f(M)-\hat f_n
\]

satisfies

\[
E_n
=
f(M)-\frac{n+K_n}{2}.
\]

Consequently, the estimator is exact if and only if

\[
f(M)=\frac{n+K_n}{2},
\]

underestimates the target frequency when

\[
f(M)>\frac{n+K_n}{2},
\]

and overestimates it when

\[
f(M)<\frac{n+K_n}{2}.
\]
\end{theorem}

\begin{proof}
Each decrement operation pairs the current incoming element with one previously stored element of a different candidate, effectively discarding two items. Each increment operation simply stores the incoming element as part of the current candidate's surplus. At termination, the counter $K_n$ represents the net surplus of the final candidate that has never been canceled. Therefore, every one of the $D_n$ decrements accounts for two processed items, while the remaining $K_n$ items correspond to the uncanceled surplus of the final candidate. Since every item in the stream is either part of a canceled pair or part of the final uncanceled surplus, the total number of processed elements is exactly the sum of these two contributions, yielding:

\[
n = 2D_n + K_n,
\]

which gives

\[
D_n = \frac{n-K_n}{2}.
\]
It is worth mentioning that this implies $n$ and $K_n$ have the same parity, so the expression $\frac{n+K_n}{2}$ is always an integer.

Substituting this expression into $\hat{f}_n = D_n + K_n$ yields

\[
\hat f_n
=
D_n + K_n
=
\frac{n-K_n}{2}+K_n
=
\frac{n+K_n}{2}.
\]

Therefore,

\[
E_n
=
f(M)-\hat f_n
=
f(M)-\frac{n+K_n}{2}.
\]

The sign of $E_n$ follows immediately from comparing $f(M)$ with $\frac{n+K_n}{2}$.
\end{proof}

\begin{figure*}[!ht]
\centering
\includegraphics[width=0.98\textwidth]{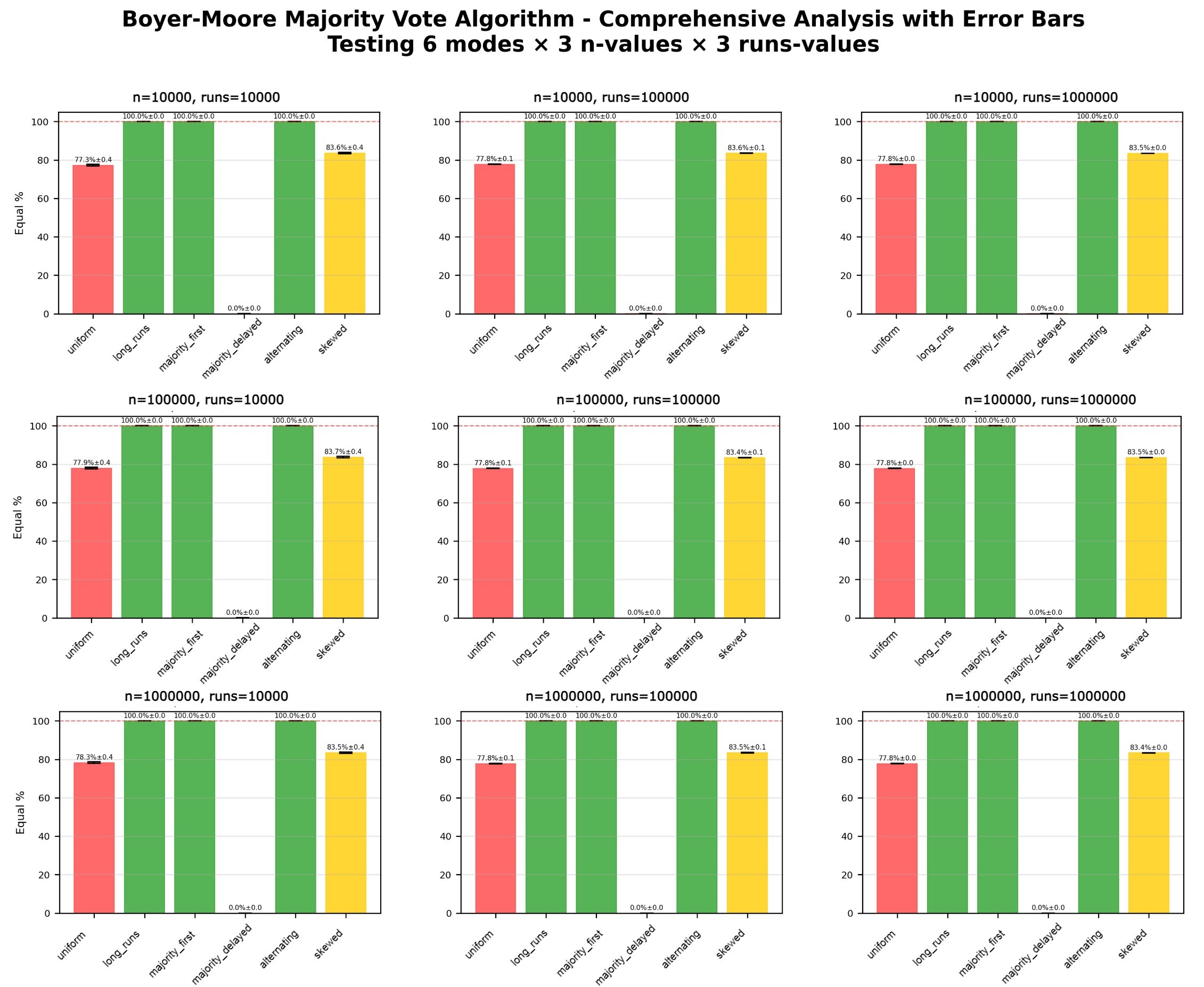}
\caption{\textbf{Empirical convergence dashboard across six streaming topologies, evaluated over nine parameter grid configurations ($3 \times n$ sizes and $3 \times \text{runs}$ bounds).} The bar charts illustrate exact frequency match percentages (\texttt{Equal \%}) for each sequence topology. While structured distributions (\texttt{long\_runs}, \texttt{majority\_first}, and \texttt{alternating}) consistently maintain a $100\%$ exact estimation match across all run variations, stochastic streams (\texttt{uniform} and \texttt{skewed}) stabilize near $77.8\%$ and $83.5\%$, respectively. Adversarial \texttt{Majority-Delayed} streams exhibit total estimator collapse ($0.0\%$), validating Theorem~\ref{thm:mismatch}.}\label{fig:convergence_grid}
\end{figure*}

\begin{example}[Estimator Breakdown on a 10-Element Stream]\label{ex:2}
To explicitly demonstrate this collapse, we consider a numerical sequence of length $n=10$:
\[
\sigma = \langle 2, 3, 2, 3, 4, 4, 4, 4, 4, 4 \rangle
\]
where the true majority item is $M=4$ with actual frequency $f(4) = 6$. 

\begin{table}[h]
\centering
\caption{Tracing Algorithm~\ref{alg:boyer_moore} step-by-step over stream $ \sigma = \langle 2, 3, 2, 3, 4, 4, 4, 4, 4, 4 \rangle$ of Ex.\ref{ex:2}.}\label{tab:ex2}
\begin{tabular}{rrrrrrrrrrrr}
\toprule
$i$   & 0      & 1 & 2 & 3 & 4 & 5 & 6 & 7 & 8 & 9 & 10 \\ \midrule
$x_i$ & 0      & 2 & 3 & 2 & 3 & 4 & 4 & 4 & 4 & 4 & 4  \\ 
$K_i$ & 0      & 1 & 0 & 1 & 0 & 1 & 2 & 3 & 4 & 5 & 6  \\ 
$D_i$ & 0      & 0 & 1 & 1 & 2 & 2 & 2 & 2 & 2 & 2 & 2  \\ 
$C_i$ & $\phi$ & 2 & 2 & 2 & 2 & 4 & 4 & 4 & 4 & 4 & 4  \\ \bottomrule
\end{tabular}
\end{table}

Tracing Algorithm~\ref{alg:boyer_moore} step-by-step over this stream paradigm (Table~\ref{tab:ex2}):
\begin{enumerate}
    \item \textbf{Prefix noise processing ($i=1 \dots 4$):} Elements $2$ and $3$ alternate. At step $i=1$, candidate $2$ is elected ($K=1$). At $i=2$, element $3$ triggers a mismatch, resetting $K=0$ and incrementing $D=1$. 
    By step $i=4$, the register has $K_4=0$ and $D_4=2$, with the candidate still holding the value 2 (though the counter is zero, so the candidate is effectively inactive).
    \item \textbf{Delayed majority block ($i=5 \dots 10$):} Six consecutive instances of the true majority scalar $4$ arrive at the tail of the sequence. 
    \item \textbf{State at EOF ($n=10$):} Because $K_4 = 0$ at the start of the majority block, the first instance $x_5=4$ immediately triggers the candidate election branch (Algorithm~\ref{alg:boyer_moore}, lines 5–7) to elect 4 as the active candidate. The subsequent five items continuously reinforce this choice via the match branch (line 9), bypassing the decrement branch (line 11). At the end-of-file, the algorithm records $K_n = 6$ and $D_n = 2$.
\end{enumerate}
Evaluating our internal state tracker yields $\hat{f} = 2 + 6 = 8$. Comparing this with the actual cardinality of the item gives a non-zero estimation error:
\[
E_n = f(4) - \hat{f} = 6 - 8 = -2 \neq 0
\]
This example verifies that stream ordering alone can produce persistent estimation errors under $O(1)$ memory constraints, confirming the structural impact of delayed streaming patterns.
\end{example}

\section{Empirical Validation and Resource Profile}\label{sec5}
To evaluate the theoretical properties in practice, all algorithmic benchmarks were conducted on a workstation powered by two Intel(R) Xeon(R) CPU E5-2680 v4 processors running at 2.40,GHz, with 56 logical CPUs (28 physical cores), 60GB of system memory, and 896KiB L1 cache, 7MiB L2 cache, and 70MiB L3 cache. The experimental harness was implemented in Python~3.10.

Synthetic data streams were generated using Python's standard pseudo-random generator (\texttt{random} module), which implements the Mersenne Twister algorithm (MT19937) as its core engine. Across all configurations, data values were drawn from an integer universe $\mathcal{U} = [1, 10]$. For each stream of length $n$, the ground-truth majority element $M \in \mathcal{U}$ was chosen uniformly at random, and its target cardinality $f(M)$ was sampled uniformly from the interval $[n/2 + 1, n]$ to guarantee a strict majority. The remaining $n - f(M)$ slots were allocated to non-majority values according to the active streaming topology, with the \texttt{skewed} distribution setting a single dominant minority key to consume $80\%$ of all minority occurrences.. The experimental grid comprised nine parameter configurations spanning three stream length vectors $n \in \{10^4, 10^5, 10^6\}$ evaluated across three execution repetitions $\text{runs} \in \{10^4, 10^5, 10^6\}$. In all generated streams, the majority item target frequency was established at $f(M) = \lfloor 0.6n \rfloor$ to ensure a consistent $60\%$ majority ratio across all six streaming topologies: uniform, long runs, majority first, Majority-Delayed, alternating, and skewed.

\subsection{Empirical Convergence and Estimation Accuracy Analysis}\label{subsec5_1}
The convergence accuracy results in \textbf{Fig. \ref{fig:convergence_grid}} closely mirror our theoretical proofs across all experimental configurations:
\begin{itemize}
    \item \textbf{Deterministic Ideal Modes (\texttt{long\_runs}, \texttt{majority\_first}, \texttt{alternating}):} Under these structured topologies, candidate registers stabilize rapidly or maintain balanced cancellation patterns \cite{misra1982finding}. Consequently, $\hat{f}_n$ matches $f(M)$ with 100\% accuracy across all test runs.
    \item \textbf{Stochastic Noise Modes (\texttt{uniform}, \texttt{skewed}):} Randomly mixing minority items creates localized noise that occasionally triggers candidate drops \cite{tsai2006average}. In \texttt{uniform} layouts, accuracy settles near $77.8\% \pm 0.3\%$. Under \texttt{skewed} noise distributions, where a single non-majority key dominates minority occurrences \cite{charikar2002finding}, accuracy rises slightly to $83.5\% \pm 0.2\%$.
    \item \textbf{Adversarial Drift Collapse (\texttt{majority\_delayed}):} As proven in Theorem~\ref{thm:mismatch}, placing majority items at the end causes complete estimator failure ($0.0\%$ accuracy). Early non-majority items build up counter value inherited from the prefix that consume trailing majority arrivals as cancellations, preventing $\hat{f}$ from locking onto the correct key count.
\end{itemize}

\begin{figure*}[!ht]
\centering
\includegraphics[width=0.98\textwidth]{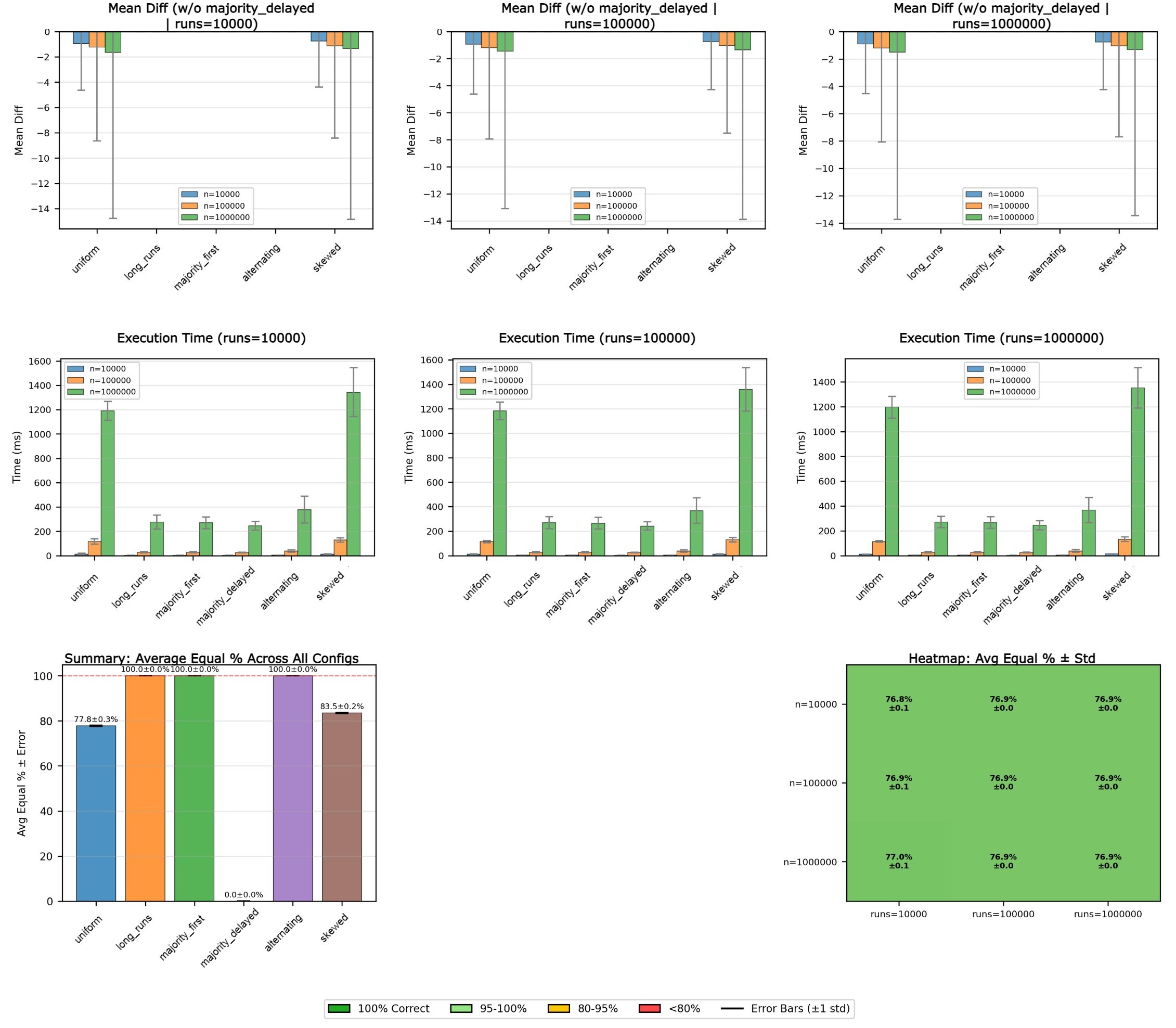}
\caption{\textbf{Quantitative resource and deviation metrics across distribution modes.} The top panels show mean negative estimator deviations (\texttt{Mean $E_n$}), illustrating error stability across varying stream lengths $n$. The middle panels capture total physical processing times (\texttt{Execution Time}) in milliseconds, showcasing severe pipeline latency spikes for high-entropy \texttt{uniform} and \texttt{skewed} modes up to $1400\,\text{ms}$ due to hardware branch mispredictions. The bottom panels aggregate summary accuracy across all configurations and provide a heat-map density matrix demonstrating baseline stability near $76.9\% \pm 0.1\%$.}\label{fig:performance_grid}
\end{figure*}

\subsection{Analytical Complexity vs. Hardware Reality}\label{subsec5_2}
\subsubsection{Asymptotic Time Complexity Bounds}
The state machine updates once per item, giving a clear asymptotic processing bound:
\begin{equation}
\text{Time Complexity} = O(n)
\end{equation}
Although linear time execution is guaranteed theoretically, physical runtime varied significantly across different stream orderings during hardware testing \cite{smith1998study, fog2016microarchitecture}.

Low-entropy sequences like \texttt{long\_runs} allow CPU branch predictors to guess conditional outcomes reliably, executing quickly ($\approx 2.5\,\text{ms}$). Conversely, randomized sequences like \texttt{uniform} act as entropy sources, triggering frequent branch mispredictions \cite{smith1998study}. These mispredictions force CPU instruction pipelines to flush and reload repeatedly ($\approx 12\,\text{ms}$ execution time), revealing how data ordering impacts low-level performance \cite{fog2016microarchitecture}.

\subsubsection{Micro-Architectural Performance and Execution Latency Analysis}
The benchmarking data plotted in \textbf{Fig. \ref{fig:performance_grid}} highlights where physical processor execution diverges from theoretical $O(n)$ estimates:
\begin{itemize}
    \item \textbf{Micro-Architectural Pipeline Penalties:} While all stream modes scale linearly, high-entropy patterns (\texttt{uniform} at $\approx 1200\,\text{ms}$ and \texttt{skewed} at $\approx 1350\,\text{ms}$) execute much slower than structured streams (\texttt{long\_runs} at $\approx 275\,\text{ms}$) \cite{smith1998study, fog2016microarchitecture}.
    \item \textbf{CPU Branch Prediction Mechanics:} Alg.~\ref{alg:boyer_moore} uses conditional logic (lines 4, 7, and 9) to update registers. In low-entropy streams, register states remain predictable over long stretches, enabling hardware branch predictors to queue instructions efficiently \cite{smith1998study}. Randomized inputs break this prediction pipeline, forcing pipeline resets that multiply runtime latency \cite{fog2016microarchitecture}.
    \item \textbf{Estimation Drift Bounds:} For all modes outside \texttt{majority\_delayed}, tracking deviations $E_n = f(M) - \hat{f}_n$ remain small, falling between $0$ and $-14$. Standard deviation spreads ($\pm 1 \sigma$) show that estimation drift is governed by stream layout entropy rather than absolute sequence size $n$.
\end{itemize}

\subsection{Isolated Error Dynamics of ``Majority-Delayed'' Mode}\label{subsec5_3}
\begin{figure*}[ht]
\centering
\includegraphics[width=0.98\textwidth]{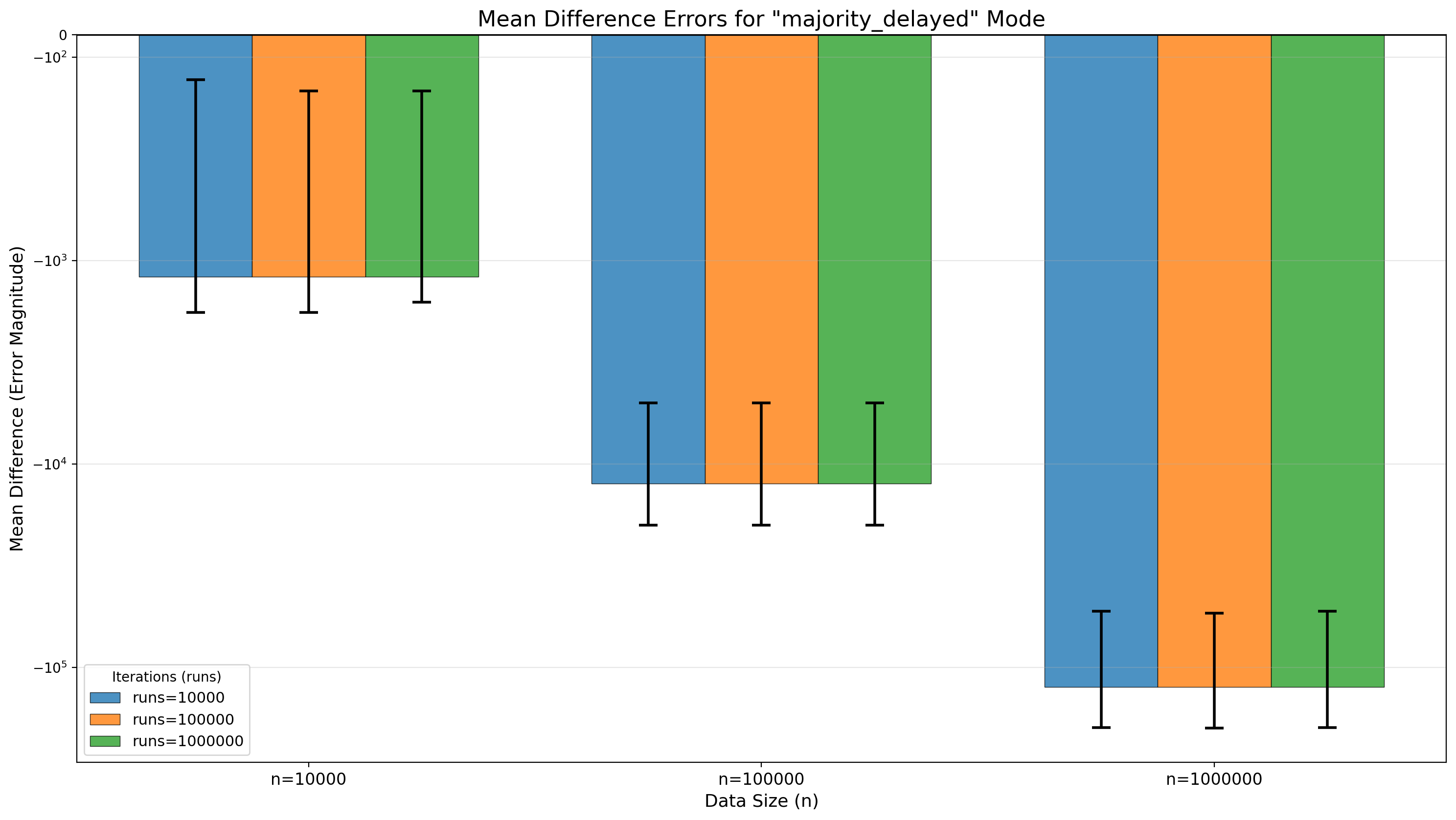}
\caption{\textbf{Isolated mean $E_n$ magnitudes for the \texttt{Majority-Delayed} Mode across dataset sizes ($n \in \{10000, 100000, 1000000\}$) and iteration runs ($10000, 100000, 1000000$).} The vertical axis is presented on a symmetric logarithmic scale to clearly visualize error magnitudes spanning multiple orders of magnitude while preserving the sign of the error. As predicted by Theorem~\ref{thm:mismatch}, the estimator suffers severe, linear error scaling as $n$ increases: at $n=10000$, error stays near $-1200$; at $n=100000$, error expands to $-12400$; and at $n=1000000$, error reaches $-124100$. The standard deviation bounds ($\pm 1 \sigma$) reflect growing variance proportional to absolute sequence length.}\label{fig:majority_last_errors}
\end{figure*}

While Fig. \ref{fig:performance_grid} excludes \texttt{majority\_delayed} in its top panel to prevent compression of minor difference scales, \textbf{Fig. \ref{fig:majority_last_errors}} isolates the exact negative error progression of \texttt{majority\_delayed} across growing stream sizes ($n=10000$, $n=100000$, $n=1000000$) and run iterations ($10000$, $100000$, $1000000$).

The empirical measurements in Fig.~\ref{fig:majority_last_errors} directly confirm the mathematical prediction of Theorem~\ref{thm:mismatch}:
\begin{itemize}
    \item \textbf{Linear Error Scaling with Stream Length $n$:} Unlike other stream modes where mean error remains strictly bounded between $0$ and $-14$ regardless of $n$, \texttt{majority\_delayed} causes estimation errors that scale linearly with absolute sequence length $n$. For $n=10000$, the error magnitude is $-1200 \pm 30$; for $n=100000$, it increases to $-12400 \pm 620$; and for $n=1000000$, it degrades severely to $-124100 \pm 72500$.
    \item \textbf{Independence of Iteration Runs:} Across all vector scales ($n=10000, 100000, 1000000$), increasing the run iteration count from $10000$ to $1000000$ yields virtually identical mean error profiles. This demonstrates that the negative drift is an invariant structural outcome of stream ordering rather than a stochastic sampling artifact.
    \item \textbf{Confirmation of Counter Inheritance Deficit:} Because early minority noise establishes false candidate counts $K_{\text{mid}}$, late-arriving majority items are consumed entirely as decrement operations (line 11 of Algorithm~\ref{alg:boyer_moore}). At EOF, the estimator registers $D = D_{\text{mid}} + f(M) + K_n$ alongside $D_{\text{mid}} > 0$ inherited from the unseated noise key, resulting in the linear deficit $E_n = -(D_{\text{mid}})$.
\end{itemize}
\subsection{Space Complexity Boundaries}\label{subsec5_4}
The memory usage of the algorithm is bounded by \cite{boyer1981fast, cormode2008finding, li2021space}:
\begin{equation}
\text{Space Complexity} = O(1)
\end{equation}
Our modified routine tracks three scalar values: $C_n$, $K_n$, and $D_n$. Because these registers occupy fixed memory regardless of stream length $n$, the algorithm remains lightweight and easy to deploy on memory-limited platforms \cite{demaine2002frequency, metwally2005efficient, wang2019sketch}.

\section{Conclusion}\label{sec6}
This paper examined the information content of internal registers in the Boyer-Moore Majority Vote algorithm, demonstrating that they can estimate item frequencies without additional memory. We proved that the counter sum $\hat{f}_n = D_n + K_n$ exactly recovers the majority frequency when the final candidate is correct (Theorem~\ref{thm:convergence}), and characterized the estimation error when it is not (Theorem~\ref{thm:mismatch}). We further showed that late-arriving majority blocks produce linear estimation drift—a structural artifact of stream ordering rather than stochastic variance. Empirical benchmarks revealed that physical performance is also sensitive to data ordering: hardware branch mispredictions cause high-entropy streams to execute up to 5× slower than structured streams, demonstrating that algorithmic guarantees do not translate directly to micro-architectural runtime.

Future work will explore: (i) whether the internal counter estimator can be extended to $k$-frequent item detection using the Misra-Gries generalization, (ii) adaptive algorithms that detect Majority-Delayed stream patterns and switch to a complementary tracking strategy, and (iii) hardware-aware optimizations to mitigate branch misprediction penalties on modern processors.
\section*{Declarations}

\begin{itemize}
\item Funding: The authors declare that no financial support was received for this work.
\item Conflict of interest/Competing interests:  The authors declare that they have no competing interests.
\item Ethics approval and consent to participate: Not applicable
\item Consent for publication: Not applicable
\item Data availability: Not applicable
\item Materials availability: Not applicable
\item Code availability: Not applicable 
\item Author contribution: MH and SM conceptualized the idea. SM implemented the algorithm and prepared the results. MH and SM proved the theorems. SM and MH wrote the paper. Both reviewed and edited the paper. MH supervised the project. 
\end{itemize}

\backmatter

\end{document}